\documentclass[runningheads]{llncs}
\usepackage{graphicx}
\begin{document}
\title{RUSLAN: RUSSIAN SPOKEN LANGUAGE CORPUS FOR SPEECH SYNTHESIS}
\titlerunning{RUSSIAN SPOKEN LANGUAGE CORPUS FOR SPEECH SYNTHESIS}

\author{Lenar Gabdrakhmanov\thanks{L. Gabdrakhmanov and R. Garaev contributed equally to this work.} \and
Rustem Garaev$^\star$ \and
Evgenii Razinkov}
\authorrunning{L. Gabdrakhmanov et al.}

\institute{Institute of Computational Mathematics and Information Technologies \\ Kazan Federal University \\ Kazan, Russia \\
\email{Evgenij.Razinkov@kpfu.ru}}
\maketitle              
\begin{abstract}
We present RUSLAN -- a new open Russian spoken language corpus for the text-to-speech task. RUSLAN contains 22200 audio samples with text annotations -- more than 31 hours of high-quality speech of one person -- being the largest annotated Russian corpus in terms of speech duration for a single speaker. We trained an end-to-end neural network for the text-to-speech task on our corpus and evaluated the quality of the synthesized speech using Mean Opinion Score test. Synthesized speech achieves 4.05 score for naturalness and 3.78 score for intelligibility on a 5-point MOS scale.
\end{abstract}
\section{Introduction}
\label{sec:intro}

Spoken language is an essential tool for human communication. In a world of AI systems and mobile computers, humans also communicate with machines. Ability to communicate with a machine using natural spoken language contributes greatly to the user experience. Two tasks should be completed in order to make it possible: speech synthesis and automatic speech recognition.

The main goal of text-to-speech systems is to generate an audio signal containing natural speech corresponding to the input text. There are several possible solutions. Speech synthesis systems based on concatenation and statistical parametrization might produce acceptable results in terms of quality but they require deeply annotated speech corpora. Providing this level of speech annotation is a time-consuming process that requires specific lexicology knowledge. Another possible drawback of this approach is language-dependent system design~\cite{taylor2009text}. 

Recent advances in deep learning resulted in significant improvements in speech synthesis task~\cite{van2016wavenet,wang2017tacotron,ping2018deep,arik2017deep,sotelo2017char2wav,shen2017natural}. Deep learning techniques excel at leveraging large amounts of training data. Thus, the quality of the synthesized speech for text-to-speech systems based on deep learning is heavily influenced by the quality and size of the speech corpus. 

Depending on the text-to-speech neural network architecture various levels of corpus annotation might be required. While WaveNet~\cite{van2016wavenet} neural network relies on extensive annotation (linguistic features,  fundamental frequency etc.), text-audio pairs are sufficient for more recent Tacotron~\cite{wang2017tacotron} end-to-end architecture. 
It is much easier to collect text-audio pairs and it would lead to larger corpora and higher quality of synthesized speech in the future.

While large open speech corpora exist for some of the most widespread languages~\cite{garofolo1993darpa,panayotov2015librispeech,honnet2017siwis} this is not the case for many other languages.

Russian is the sixth most widespread language in the world by the number of native speakers being spoken by approximately 154 million people worldwide~\cite{Ethnologue21}. However, publicly available and annotated speech corpora in Russian are not sufficient. 
Availability of large amounts of annotated speech is crucial for the research community both in speech synthesis and recognition. 

Amount of speech for a single speaker is an important factor for end-to-end neural speech synthesis.
There are few Russian speech corpora: ~\cite{voxforge_ru,festvox_ru} are public and ~\cite{skrelin2010corpres,kachkovskaia2016coruss} are proprietary, but the amount of speech for a single speaker in these corpora is less than 7 hours.
Up until now, there was only one open-source Russian language speech corpus exceeding 7 hours in audio duration for a single speaker namely M-AILABS~\cite{MAILabs}, containing about 20 hours of speech for a single speaker at most.
In this work, we try to facilitate research in speech synthesis in Russian by providing large publicly available annotated speech corpus.

Our contributions are as follows:
\begin{itemize}
    \item We collected the largest annotated speech corpus in the Russian language for a single speaker -- RUSLAN (RUSsian spoken LANguage corpus) and made it publicly available. Our corpus is 50\% larger in terms of audio duration in comparison with the second largest corpus in the Russian language for a single speaker to date~\cite{MAILabs}.

        Speech corpus is publicly available under Creative Commons BY-NC-SA 4.0 license at https://ruslan-corpus.github.io.
    \item We trained text-to-speech neural network on RUSLAN and evaluated the quality of the synthesized speech using Mean Opinion Score with 50 participating native speakers as respondents~\footnote{Audio samples from corpus and examples of synthesized speech can be found at https://ruslan-corpus.github.io}. 
    \item We propose several improvements for Tacotron text-to-speech end-to-end neural network that allow us to achieve comparable speech quality in fewer training iterations. 
\end{itemize}

\section{Speech Corpus}
\label{sec:speech_corpus}
\begin{figure*}
\begin{minipage}[b]{1.0\linewidth}
  \centering
  \centerline{\includegraphics[width=\textwidth]{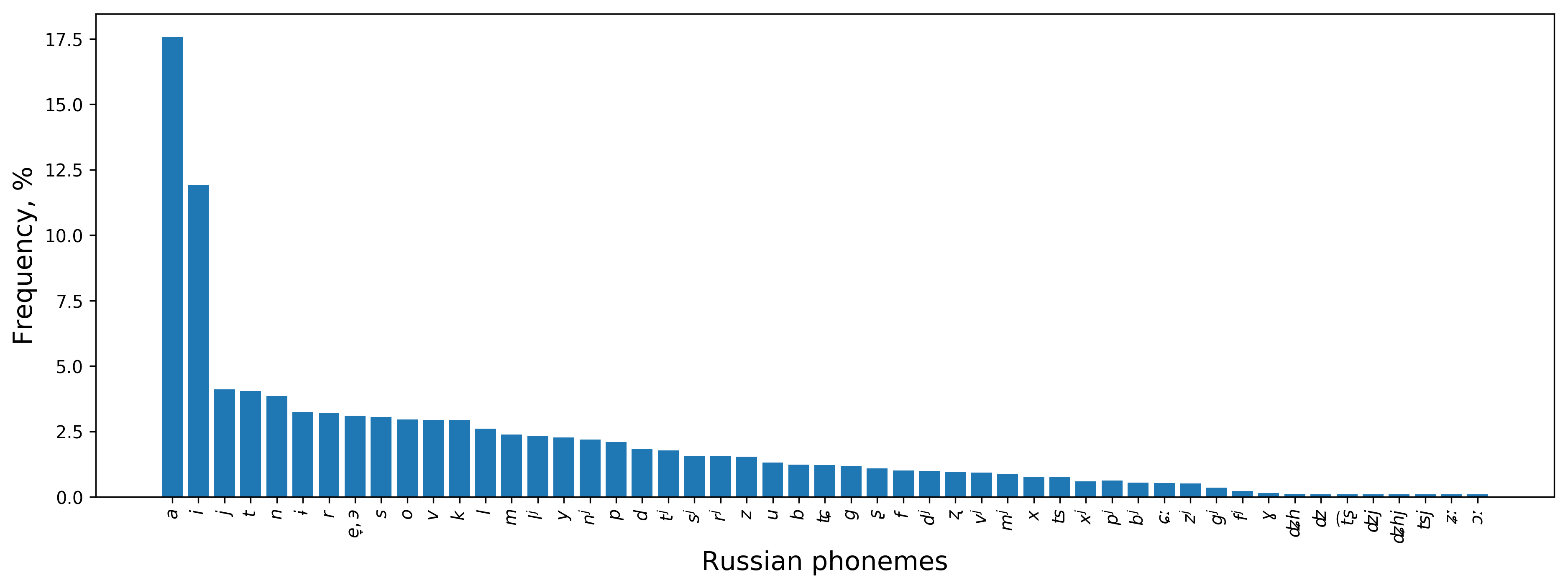}}
\end{minipage}
\caption{Distribution of the Russian phonemes in the corpus.}
\label{fig:phonemes_dist}
\end{figure*}

In this section, we describe RUSLAN speech corpus. The amount of annotated speech for a single speaker is a key feature of text-to-speech corpora. Therefore, we focused on maximizing the amount of high-quality speech recording for a single speaker. Speaker is a 23 years old male who is a native Russian speaker. 
The pronunciation is clear and intelligible.
The style of the text is narrative, the speech is neutral.
Corpus contains 22200 training samples. Each training sample is a text-audio pair, where the text is a phrase or a sentence -- an excerpt from works of Russian and American writer Sergei Dovlatov. The number of words in each training sample varies from 1 to 111 with an average of 12. 
The Russian language consists of 33 letters (10 vowels and 23 consonants). The frequency distribution for the phonemes are provided in Figure~\ref{fig:phonemes_dist}. Phonemic transcription was performed as suggested in~\cite{yakovenko2018algorithms}.

\subsection{Text preprocessing}
  Text for each training sample was preprocessed in the following way: 
\begin{itemize}
      \item All numbers and dates were manually replaced by their textual representation. 
      \item Acronyms were manually substituted with their expanded forms.
      \item All symbols except for Russian letters and punctuation marks were automatically deleted.
\end{itemize}

\subsection{Recording process}
Audio samples were recorded in a quiet and noise-protected room using noise-reduction hardware. Each sample was recorded separately with a sampling frequency of 44.1 kHz and 16 bit linear PCM and saved in WAV format. Leading and trailing silent parts were deleted from each audio sample. All text-audio pairs were additionally verified in order to avoid annotation errors. 
The signal-to-noise ratio is approximately equal to 90 dB. Corpus statistics are presented in table \ref{vo_stats}.

\begin{table}[ht]
\caption{RUSLAN corpus statistics}
\centering
\begin{tabular}{|l|l|}
\hline
Total duration                           & 31:32:55 \\
\hline
Total number of samples                  & 22200    \\
\hline
Total symbols                            & 1472377  \\
\hline 
Total words                              & 267053   \\
\hline
Unique words                          & 52703   \\
\hline
Min sample duration                   & 0.61 sec   \\
\hline
Max sample duration                   & 50.71 sec  \\
\hline
Min number of symbols in one sample   & 9   \\
\hline
Max number of symbols in one sample   & 596   \\
\hline
Min number of words in sample         & 1   \\
\hline 
Max number of words in sample        & 111   \\
\hline
\end{tabular}
\label{vo_stats}
\end{table}

\begin{figure}[ht]
\begin{minipage}[b]{0.48\linewidth}
  \centering
  \centerline{\includegraphics[width=4.0cm]{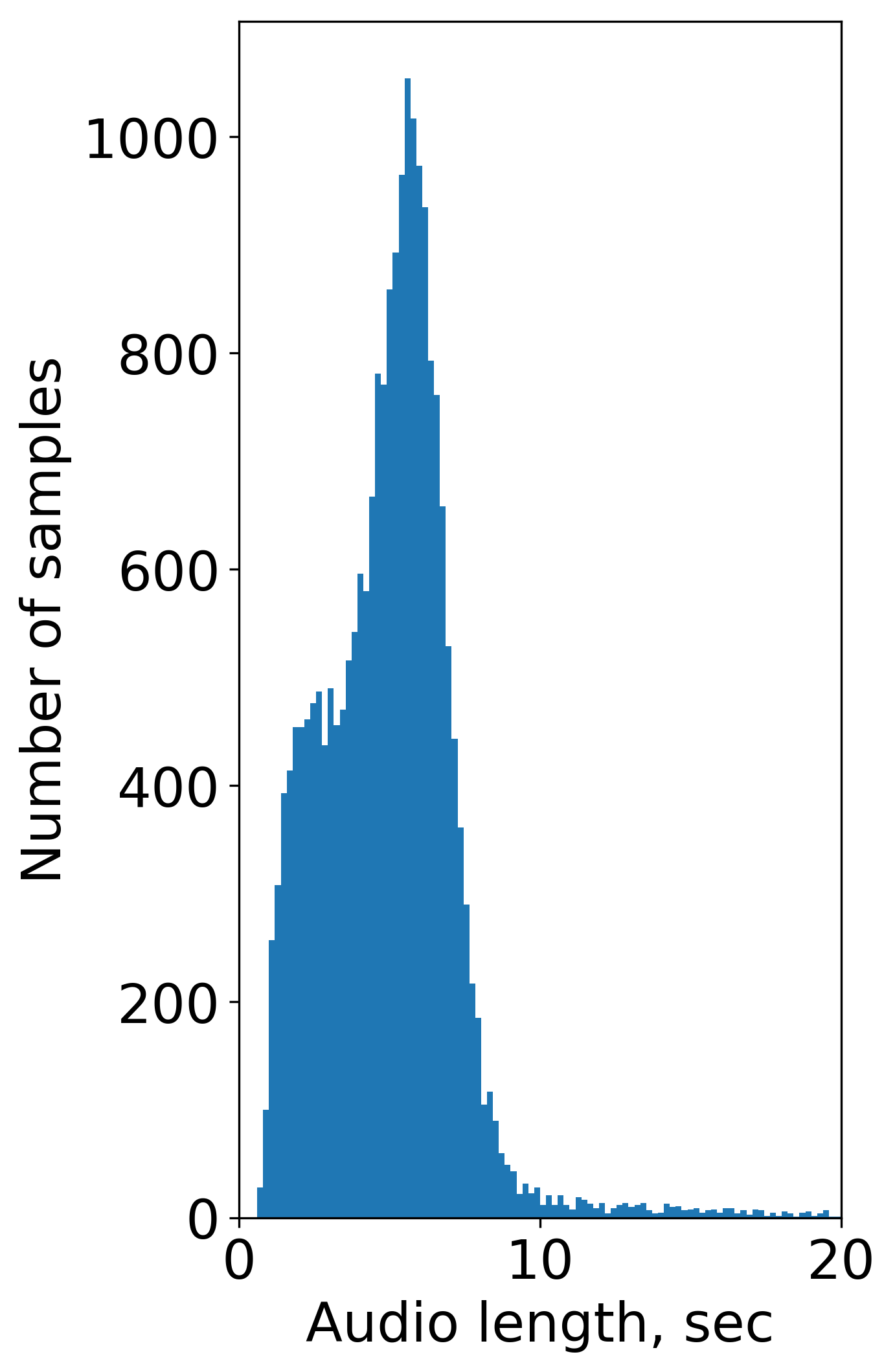}}
  \centerline{(a) Duration of samples}\medskip
\end{minipage}
\hfill
\begin{minipage}[b]{0.48\linewidth}
  \centering
  \centerline{\includegraphics[width=4.0cm]{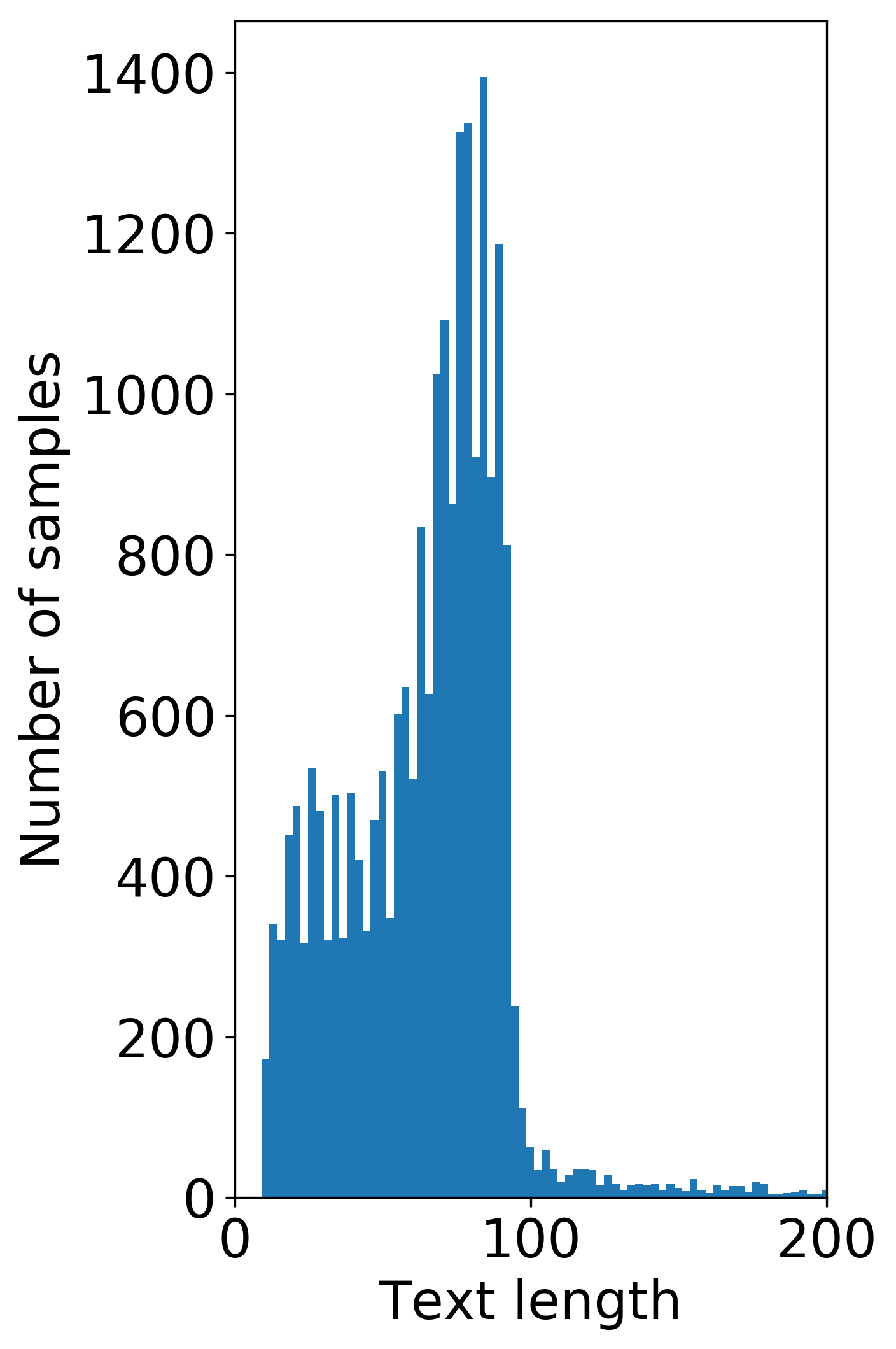}}
  \centerline{(b) Number of symbols}\medskip
\end{minipage}
\caption{Histograms (a) of the duration of samples, (b) of the number of symbols.}
\label{fig:hists}
\end{figure}

Figure~\ref{fig:hists} shows the ratio of the lengths and ratio of the number of symbols per sample for the whole corpus.

\section{Neural network for speech synthesis}
\label{sec:nn_for_ss}

\begin{figure*}
\begin{minipage}[b]{1.0\linewidth}
  \centering
  \centerline{\includegraphics[width=\textwidth]{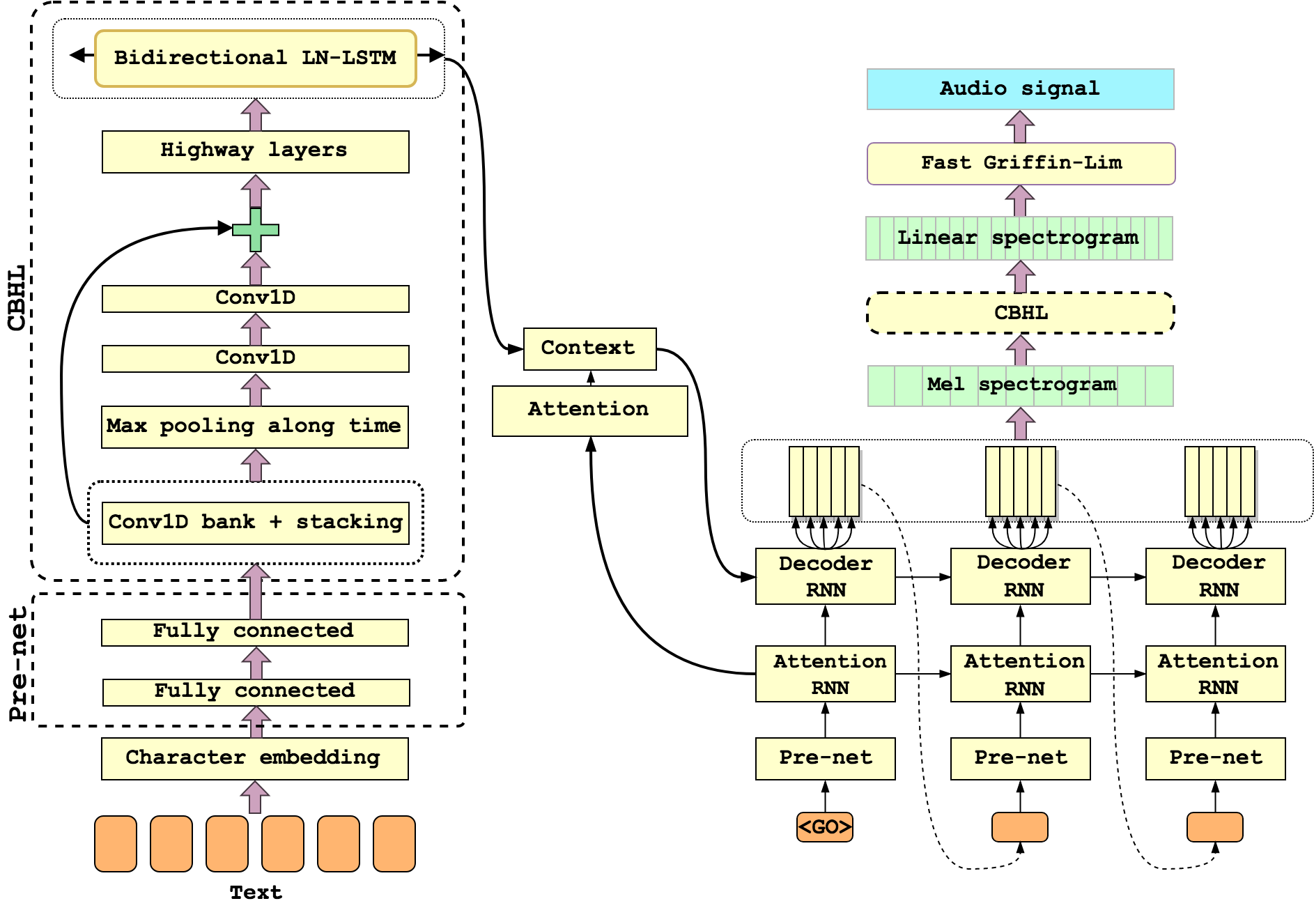}}
\end{minipage}
\caption{Model architecture.}
\label{fig:our_model}
\end{figure*}

In order to evaluate sufficiency and completeness of RUSLAN for Russian speech synthesis, we train a neural network for the text-to-speech task. In this section, we describe our neural network that is heavily based on Tacotron architecture~\cite{wang2017tacotron} with few changes that improve convergence and synthesized speech quality which we discuss below.



We employ end-to-end trainable encoder-decoder deep neural network architecture that receives text as an input and produces a linear spectrogram. This spectrogram is later used for waveform reconstruction. Model architecture is illustrated in Figure~\ref{fig:our_model}. We describe the encoder, decoder and audio reconstruction procedure below.


\subsection{Neural network architecture}

The input of the model is a text where each distinct character is represented as a trainable 256-dimensional character embedding vector. 
Thus, the lookup table has a shape of $78 \times 256$ since we use only 78 characters: Russian capital and lowercase letters, space and punctuation marks -- \{$',-().:;!?$\}.

The model encoder consists of two parts: pre-net of two fully connected layers with dropout~\cite{srivastava2014dropout} and CBHL module which is a slight modification of Tacotron CBHG module. Our main and only modification here is a replacement of GRU with layer normalized LSTM (LN-LSTM) in bidirectional RNN for faster convergence~\cite{lei2016layer}.

Decoder consists of two parts to predict mel-frequency cepstral coefficients and linear spectrogram respectively. The first part includes pre-net, Attention RNN and Decoder RNN. We have replaced GRU with LN-LSTM in both Attention RNN~\cite{bahdanau2014neural} and Decoder RNN parts in contrast with the original Tacotron model. In Tacotron model the second part of the decoder is post-processing CBHG module, but we again replace it with our CBHL.


\subsubsection{Loss function}
Since Decoder RNN predicts MFCCs and post-processing CBHL module predicts linear spectrogram, we employ two different loss functions.

Target values for Decoder RNN are 80-band MFCCs:
\begin{equation}
     Loss_{mel} = \frac{1}{N} \sum_{i}^N |\mathbf{t}^{mel}_i - \mathbf{y}_{mel}(text_i)|_1,
\end{equation}
where $N$ is the number of samples in the training set, $text_i$ is the $i$-th text from the corpus, $\mathbf{t}^{mel}_{i}$ is ground truth mel-frequency cepstral coefficients for $text_i$, $\mathbf{y}^{mel}(text_i)$ is MFCCs predicted by Decoder RNN of the neural network given $text_i$ as an input.

Loss function for post-processing CBHL module:
\begin{equation}
Loss_{lin} = \frac{1}{N} \sum_{i}^N |\mathbf{t}^{lin}_i - \mathbf{y}_{lin}(text_i)|_1,
\end{equation}
where $t^{lin}_j$ is ground truth linear-spectral coefficients for $text_i$, $\mathbf{y}_{lin}(text_i)$ is linear-spectral coefficients predicted by post-processing CBHL module of the neural network given $text_i$ as an input.

The overall loss function of the neural network is computed as follows:
\begin{equation}
Loss = Loss_{mel} + Loss_{lin}.
\end{equation}

\subsubsection{Signal reconstruction}
In contrast to Tacotron we employ fast Griffin-Lim~\cite{perraudin2013fast} algorithm to reconstruct an audio signal from magnitude-only values of the linear spectrogram. 

The signal is being recovered iteratively, we stop the process after 300 iterations. Optimization speed $\alpha$ was set to 0.99.
 
\subsection{Training}
Text from each text-audio pair from RUSLAN corpus was used as a training sample and corresponding audio was used to obtain target variables, MFCCs and a linear spectrogram.  

Our model implementation had been training for 300K iterations with a batch size of 8. The training was performed on a single GTX 1060 with 6 Gb of onboard memory. We used Adam optimizer~\cite{kingma2014adam} with exponential learning rate decay.

\section{Evaluation}
\label{sec:eval}

Synthesized speech is intelligible, natural and close to human speech. The described model shows good results even on a low amount of steps. 


\subsection{Mean Opinion Score}
Mean Opinion Score (MOS) is the most frequently used method of a subjective measure of speech quality. MOS is used to evaluate methods of signal processing, including speech synthesis. The respondents rate the speech quality on a five-point scale. Score 1 corresponds to bad quality, score 5 corresponds to excellent quality. The final rating of the signal in question is calculated as the mean over rating scores from all respondents. This method is recommended by ITU and IEEE for the quality estimation of the synthesized speech~\cite{rothauser1969ieee,itu1990}.
The scale of MOS scores is presented in table~\ref{mos_def}:

\begin{table}[ht]
\caption{MOS scale}
\centering
\begin{tabular}{|l|l|l|}
\hline
\multicolumn{1}{|c|}{Score} & \multicolumn{1}{c|}{Quality} & \multicolumn{1}{c|}{Distortions}\\ \hline
5 & Excellent & Imperceptible\\ \hline
4 & Good & Tangible, but non-irritating\\ \hline
3 & Fair & Sensible and slightly annoying\\ \hline
2 & Poor & Annoying\\ \hline
1 & Bad & Annoying and unpleasant\\ \hline
\end{tabular}
\label{mos_def}
\end{table}

In our work, we rate speech intelligibility and naturalness. The respondents were allowed to listen to the samples on their own equipment in an uncontrolled environment. The scores produced by this method remain very close to those received in a controlled environment as it was mentioned in the work~\cite{ribeiro2011crowdmos}. 50 respondents participated in the synthesis speech evaluation survey at the age of 20-40: 40\% of them were female and 60\% were male. 

Twenty audio samples consisting of 11 samples of synthesized speech and 9 samples of recorded speech from the corpus were blindly presented to the respondents. Each respondent got acquainted with the survey rules in advance.
Table \ref{table_mos} shows naturalness and intelligibility scores for the synthesized speech and original recordings from the corpus. 
In contrast to the similarity of scores achieved by real speech, the difference between naturalness and intelligibility scores for synthesized speech might be explained by the fact that respondents can understand all words clearly, although there are still some artefacts in audio due to use of the Fast Griffin-Lim algorithm.

\begin{table}[ht]
\caption{Naturalness and intelligibility scores}
\centering
\begin{tabular}{|l|l|l|}
\hline
\multicolumn{1}{|c|}{Type} & \multicolumn{1}{c|}{Naturalness} & \multicolumn{1}{c|}{Intelligibility}\\ \hline
Real speech & 4.83 & 4.87\\ \hline
Synthesized speech & 3.78 & 4.05\\ \hline
\end{tabular}
\label{table_mos}
\end{table}

We also evaluated the quality of the speech synthesized by our implementation of the reference Tacotron neural network that had been training on RUSLAN for 300K iterations. Speech synthesized by Tacotron achieved 3.12 MOS score for intelligibility and 2.17 MOS score for naturalness. 
It should be noted, however, that in the original paper Tacotron neural network was trained for 2 million iterations~\cite{wang2017tacotron}.

\section{Conclusion}
\label{sec:conclusion}
We present RUSLAN spoken language corpus -- the largest Russian open speech corpus for a single speaker for the text-to-speech task. It consists of 22200 text-audio pairs with the total audio duration being 31 hours 32 minutes and exceeds the second largest Russian corpus for a single speaker by 50\%. We evaluate the sufficiency and the completeness of our corpus by training an end-to-end text-to-speech neural network on RUSLAN. Our model achieves $4.05$ for intelligibility and $3.78$ for naturalness on a 5-point MOS scale.

\bibliographystyle{splncs04}
\bibliography{ruslan_paper}

%
%

\end{document}